\documentclass[aps,prb,twocolumn,psfig,showpacs]{revtex4}
\usepackage{amsfonts}
\usepackage{amsmath}
\usepackage{graphicx}
\usepackage{bm}
\usepackage{amssymb}
\usepackage{times}
\usepackage{dcolumn}
\usepackage{cases}
\usepackage{txfonts}
\usepackage{color}

\RequirePackage[hyperindex,colorlinks,bookmarksnumbered,plainpages=true]{hyperref}
\hypersetup{colorlinks,linkcolor=blue,urlcolor=blue,citecolor=blue}
\usepackage{hyperref}

\usepackage{ulem}
\renewcommand{\emph}[1]{\textit{#1}}

\definecolor{darkgreen}{rgb}{0,0.5,0}
\definecolor{darkblue}{rgb}{0,0,0.5}
\definecolor{darkred}{rgb}{.7,0,0}
\definecolor{purple}{rgb}{0.35,0,0.35}
\definecolor{orange}{rgb}{1,0.5,0}
\definecolor{grey}{rgb}{.6,.6,.6}

\begin{document}

\title{Identifying Symmetry-Protected Topological Order by Entanglement Entropy}
\author{Wei Li, Andreas Weichselbaum, and Jan von Delft}

\affiliation{Physics Department, Arnold Sommerfeld Center for Theoretical Physics, and Center for NanoScience, Ludwig-Maximilians-Universit\"at, 80333 Munich, Germany}

\begin{abstract}
According to the classification using projective representations of the SO(3) group, there exist two topologically distinct gapped phases in spin-1 chains. The symmetry-protected topological (SPT) phase possesses half-integer projective representations of the SO(3) group, while the trivial phase possesses integer linear representations. In the present work, we implement non-Abelian symmetries in the density matrix renormalization group (DMRG) method, allowing us to keep track of (and also control) the virtual bond representations, and to readily distinguish the SPT phase from the trivial one by evaluating the multiplet entanglement spectrum. In particular, using the entropies $S^I$ ($S^H$) of integer (half-integer) representations, we can define an entanglement gap $G = S^I - S^H$, which equals 1 in the SPT phase, and $-1$ in the trivial phase. As application of our proposal, we study the spin-1 models on various 1D and quasi-1D lattices, including the bilinear-biquadratic model on the single chain, and the Heisenberg model on a two-leg ladder and a three-leg tube. Among others, we confirm the existence of an SPT phase in the spin-1 tube model, and reveal that the phase transition between the SPT and the trivial phase is a continuous one. The transition point is found to be critical, with conformal central charge $c=3$ determined by fits to the block entanglement entropy.
\end{abstract}

\maketitle

\section{introduction}
Symmetry-protected topological (SPT) phases have attracted enormous research interest recently. \cite{Wen_1, Wen_2, Wen_3, Wen_4, Meng, Burnell, Cirac, Pollmann_1, Pollmann_2, Pollmann_3, Haegeman, Thomas} Among the interesting models exhibiting SPT order, a remarkable example is the spin-1 chain. The generic spin-1 bilinear-biquadratic (BLBQ) model can be written down as
\begin{equation}
H_{blbq} = J \sum_{<i,j>} [\cos(\theta) S_i S_j + \sin(\theta) (S_i S_j)^2],
\label{eq-bbq}
\end{equation}
where the coupling $J=1$ sets the energy scale, and $\theta$ is a tunable parameter. The phase diagram of the spin-1 BLBQ model with respect to various $\theta$'s is well known (except for a subtlety in the thin region near $\theta=-5/4 \pi$).\cite{Lauchli, Tu} When $-\pi/4 < \theta < \pi/4$, the system is in the Haldane phase. \cite{Haldane} Although this phase has been intensively studied, it has been realized to be an SPT phase only very recently.\cite{Wen_1, Cirac, Pollmann_1} At $\theta=\arctan(1/3)$, an exactly solvable point within the Haldane phase,  the ground state is termed AKLT state,\cite{Affleck} which can be exactly expressed as a matrix product state (MPS) with bond dimension $D=2$. The Haldane phase has a nonzero spin gap, called Haldane gap, which can be interpreted in terms of spinon confinement. \cite{Rachel} No local order parameter can be found to distinguish the Haldane phase from a trivial gapped phase, nevertheless, there exists a nonlocal string order parameter (SOP), \cite{Rommelse}
\begin{equation}
O_{\alpha} = - \lim_{j-i \to \infty} [S_i^{\alpha} \exp(i \pi \sum_{i<l<j} S_l^{\alpha}) S_j^{\alpha}], 
\label{eq-SOP}
\end{equation}
where $\alpha=x, z$. This string order parameter characterizes the topological order in the Haldane phase. Further studies show that the string order parameter $O_{x,z}$ can be transformed to two ordinary ferromagnetic order parameters through a nonlocal unitary transformation $U_{KT} = \prod_{k < l} \exp(i \pi S_k^z S_l^x)$. Therefore, a nonzero string order parameter actually reveals a hidden $Z_2 \times Z_2$ symmetry breaking. \cite{KT, Oshikawa}

\begin{figure}[tbp]
\includegraphics[angle=0,width=1\linewidth]{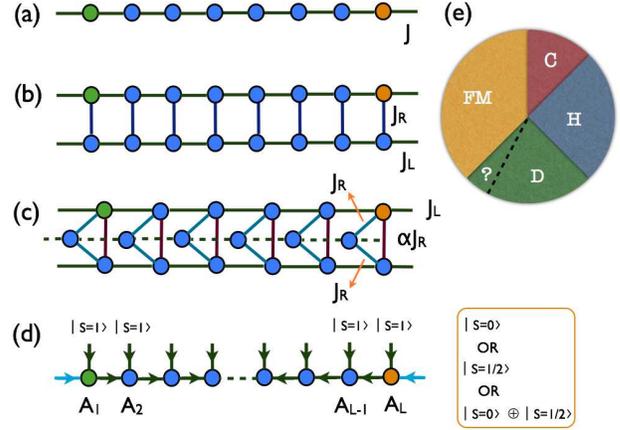}
\caption{(Color online) (a) Spin chain with coupling constant $J$. (b) Two-leg spin ladder model with $J_L$ and $J_R$ for couplings along chain and rung directions, respectively. (c) Three-leg spin tube model, $J_L$ is the coupling along the leg. Each isosceles triangle contains two kinds of couplings, $J_R$ for the two equal sides and $\alpha J_R$ for the third. (d) The SU(2)-invariant matrix product state describing ground state of spin-1 chain (a), ladder (b) or tube (c). $|S=n\rangle$ represents a multiplet with quantum number $S=n$. For the spin-1 model, each local space is a $|S=1\rangle$ triplet. The input bond multiplets on both open ends can be controlled, and three common choices are shown in (d). (e) shows the phase diagram of the spin-1 BLBQ chain,\cite{Lauchli, Tu} H, C, FM, and D stand for Haldane, critical, ferro-magnetic, and dimerized phases, respectively. There is a narrow region near $\theta \approx -3/4 \pi$ with possible spin nematic order, whose existence still remains debatable.} \label{fig-sketch}
\end{figure}

The Haldane phase is protected by several global symmetries. According to the valence bond solid (VBS) picture, the gapped Haldane phase only possesses short-range entanglement, hence it is not an intrinsic topological phase.\cite{Wen_2} Its nontrivial topological properties are protected by parity symmetry, time reversal symmetry, and $Z_2 \times Z_2$ rotational symmetry around the $x$ and $z$ axes.\cite{Wen_1, Pollmann_1} The Haldane phase can not be adiabatically connected to the trivial one as long as one of the above symmetries is preserved along the path; instead, a quantum phase transition (QPT) must occur along the way. As is well known, the Landau paradigm classifies the various symmetry-breaking phases according to symmetry groups.\cite{Landau, Landau_2} Nevertheless, the existence of a QPT between SPT phases and trivial ones shows that gapped phases without symmetry breaking in one dimension (and also in higher dimensions) can still be distinct and classified by the group cohomology of symmetries.\cite{Wen_2, Wen_3, Cirac}

To be specific, we consider the gapped phases in \mbox{spin-1} SO(3) Heisenberg chains, which can be generally classified by different projective representations of the rotational SO(3) group, i.e., the corresponding group cohomology $H^2(\rm{SO(3), U(1))=Z_2}$. \mbox{Integer-spin} (even) representations of SU(2) are linear representations of SO(3); half-integer (odd) representation, which involve an additional minus sign after $2 \pi$ rotations (owing to the SU(2) double covering, SO(3) = SU(2)/$\rm{Z}_2$) are projective representations of SO(3). Based on this observation, the classification theory states: there are two distinct gapped phases in spin-1 chains corresponding to two different kinds of representations of SO(3), linear and projective. They correspond to the trivial phase and the Haldane phase, respectively. \cite{Wen_2, Cirac}

It has recently been discovered that these two phases also differ strikingly in the structure of their entanglement spectra. The entanglement spectrum consists of the eigenvalues of the entanglement Hamiltonian $H_E = -\log(\rho)$, where $\rho$ is the reduced density matrix of a subsystem.\cite{Li} The entanglement spectrum of the bulk has an intimate relationship with the  real excitation spectrum on the boundary.\cite{Cirac2} Closely related with the group cohomology classification, an interesting feature has been found: For the spin-1 chain, the entanglement spectrum is found to show at least two-fold degeneracy for the Haldane phase, while it is generally non-degenerate for the trivial phase.\cite{Pollmann_1} The occurrence of the two-fold degeneracy can be used to numerically identify the Haldane phase. 

Actually, this degeneracy in the entanglement spectrum is a signature of the appearance of half-integer-spin multiplets in the MPS geometric bond, which support projective representations of the SO(3) group. Take the AKLT point as an intuitive example: according to the construction of the AKLT state, each local spin-1 is decomposed into two spin-1/2 ancillas. The AKLT state can be exactly expressed as MPS with bond dimension 2, hence only one $|S=1/2\rangle$ doublet appears on each of its geometric bonds, and the entanglement spectrum is two-fold degenerate. For other generic states in the Haldane phase, the multiplets on the geometric bonds are generally $S$=half-integer, which leads to at least two-fold degeneracy and supports projective representations. This key feature can be used to differentiate the SPT phase from the trivial one, the latter instead has integer bond multiplets that support linear representations. Therefore, if we could \emph{directly identify the virtual spins} on the geometric bonds of the MPS, it would be straightforward to see whether the representation is projective or linear, and thus to identify the SPT or trivial phase.

One powerful numerical method for solving 1D quantum spin models is the density matrix renormalization group (DMRG).\cite{White, Schollwoeck} In order to further improve its efficiency and stability, Abelian and non-Abelian symmetries have been implemented in the DMRG algorithm.\cite{McCulloch} In particular, the \mbox{state-of-the-art} SU(2) DMRG technique enables us to identify the spin of the multiplets on the virtual bonds. Note, though, that if open boundary conditions are adopted for SU(2) DMRG, because only integer-spin sectors are allowed by adding spin-1's together, the renormalized bases on the virtual bonds are automatically integer multiplets, i.e., linear representations of SO(3). This would imply the absence of two-fold degeneracy within each multiplet (every multiplet contains odd number of individual states) even in Haldane phase, which seems paradoxical. 

To solve this problem, we propose a protocol algorithm in this paper which automatically determines the proper bond representations. In addition, by defining and calculating the integer and half-integer entanglement entropies, we elucidate why this protocol algorithm works, and obtain a simple criterion for identifying the SPT phase. We test these ideas in three spin-1 lattice models, and show that they succeed in telling the SPT phase from the trivial one. 

The paper is organized as follows. In Sec. II, the \mbox{SU(2)-invariant} MPS and related DMRG algorithms are briefly introduced. In Secs. III-V, we show that the entanglement entropies can be used to identify the SPT phase, by studying three examples including the single spin-1 chain, 2-leg ladder and 3-leg tube models. In particular, in the spin-1 tube model, the transition between the SPT and trivial phases is verified to be a continuous QPT. Sec VI offers a summary.

\begin{figure}[tbp]
\includegraphics[angle=0,width=1.0\linewidth]{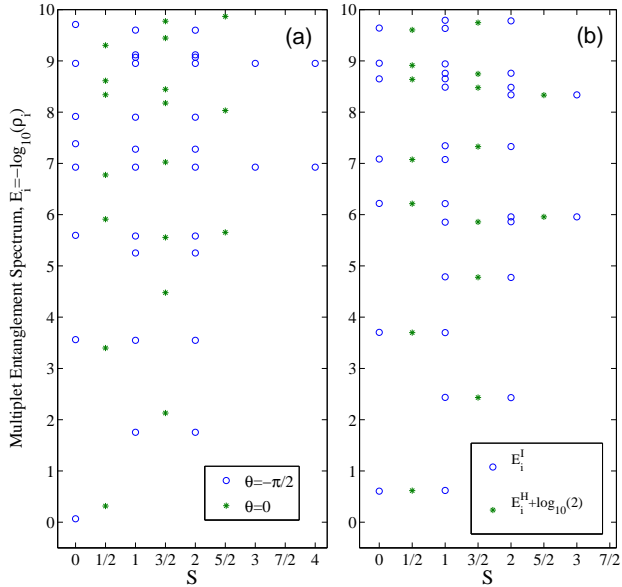}
\caption{(Color online) (a) The multiplet entanglement spectrum of a spin-1 BLBQ chain, calculated using the protocol algorithm. Multiplets $|S=0\rangle \oplus |S=1/2 \rangle$ are put on the end bonds, and the converged spectra are obtained after several DMRG sweeps. $\theta=0$ (asterisks) is in the Haldane phase, with half-integer-spin bond multiplets; $\theta=-\pi/2$ (circles) is in the trivial phase, with integer-spin bond multiplets. Every data point represents a multiplet (not as usual a single state within a multiplet). Therefore, a multiplet with symmetry label $S$ corresponds to $2S+1$ degenerate states. (b) Multiplet spectrum of integer and half-integer representations for the Haldane phase calculated for $\theta=0$, and using $|S=1/2\rangle$ (asterisks) or $|S=0\rangle$ (circles) respectively, on the end bonds. The half-integer spectrum has been shifted by $\log(2)$, in order to reveal the one-to-one correspondence between each multiplet in the half-integer spectrum and a pair of degenerate multiplets in the integer spectrum.} \label{fig-ent-spec}
\end{figure}

\section{SU(2) invariant Matrix Product States and Multiplet Entanglement Spectrum}
The variational MPS ground state of 1D Heisenberg systems with Hamiltonians like Eq.~(\ref{eq-bbq}) can be written in an SU(2)-invariant form. Corresponding bond spaces are factorized into two parts,\cite{Weich}
\begin{equation}
| \tilde{Q} \tilde{n}; \tilde{Q}_z \rangle = \sum_{Q n, Q_z} \sum_{q l,q_z} (A_{Q,\tilde{Q}}^q)_{n, \tilde{n}}^{l}(C_{Q, \tilde{Q}}^q)_{Q_z, \tilde{Q}_z}^{q_z} |Q n; Q_z\rangle |q l; q_z\rangle,
\end{equation}
where $Qn$ (and $\tilde{Q}\tilde{n}$, $ql$) are composite multiplet indices. $Q$ specifies the symmetry sector, $n$ distinguishes different multiplets with the same $Q$, and $Q_z$ ($\tilde{Q_z}$, $q_z$) labels the individual states within a given multiplet in symmetry sector $Q$ ($\tilde{Q}$, $q$).

The $A$-tensors can be regarded as physical tensors which combine the input multiplets $(Q n)$ with the local space $(q l)$, and transform (and possibly truncate) them into the output multiplets $(\tilde{Q} \tilde{n})$; the $C$-tensors are the Clebsch-Gordan coefficients (CGC) which take care of the underlying mathematical symmetry structure. The tensor product of physical tensor $A$ (reduced multiplet space) and its related mathematical tensor $C$ (CGC space) has been called the QSpace,\cite{Weich} which is a generic representation used in practice to describe all symmetry-related tensors. \cite{Weich} The QSpace is a very useful concept not only for MPS wavefunctions, but also for calculating the matrix elements of irreducible tensor operators, which can be treated in the same framework according to the Wigner-Eckart theorem.

By implementing the QSpace in our DMRG code, we need to determine only the physical $A$-tensors variationally as in plain DMRG, while the underlying CGC space ($C$-tensors) are fully determined by symmetry. The $A$-tensors manipulate multiplets $(Q n)$ only on the reduced multiplet level, which leads to a large gain in numerical efficiency. In this work, by adopting the SU(2)-invariant MPS, we are able to keep track of the quantum numbers $S$ of the bond multiplets, and hence to distinguish the SPT phase and the trivial phase straightforwardly.

Given an SU(2)-invariant MPS, it is natural to consider its \textit{multiplet entanglement spectrum}, defined of multiplets, rather than individual states. To be explicit, we note that any SU(2)-invariant MPS can be written in the following form:
\begin{eqnarray}
&&|\psi \rangle = \sum_{\{q_i^z\}} {\rm{Tr}} [(A^{q_1}_{Q_1, Q_2})^{l_1}_{n_1,n_2} (\Lambda_{Q_2})_{n_2} ... (\Lambda_{Q_{L-1}})_{n_{L-1}} (A^{q_{L}}_{Q_{L-1}, Q_{L}})^{l_L}_{n_{L-1}, n_{L}} \notag \\
&&(C^{q_1}_{Q_1, Q_2})^{q_1^z}_{Q_1^z, Q_2^z} (\lambda_{Q_2})_{Q_2^z} ... (\lambda_{Q_{L-1}})_{Q_{L-1}^z} (C^{q_L}_{Q_{L-1}, Q_L})^{q_L^z}_{Q_{L-1}^z, Q_L^z} ] |q_1^z ... q_L^z \rangle. \notag \\
\label{eq-su2-mps}
\end{eqnarray}
The trace includes all the quantum labels $(Q_i n_i,Q_i^z)$, while $q_i, l_i$ all equal 1 in the present spin-1 case. Eq.~(\ref{eq-su2-mps}) is an SU(2)-invariant version of Eq.~(4) in Ref.~\onlinecite{Pollmann_1}. The difference is that the conventional MPS matrix $A$ is represented in the factorized form of a direct product, i.e., $A_{Q_{i-1}, Q_i}^{q_i} \otimes C_{Q_{i-1}, Q_i}^{q_i}$. In Eq.~\ref{eq-su2-mps} above, we have assumed the canonical MPS forms in both the reduced multiplet space and the CGC space. Notice that since the $C$-matrices store CGC's, they automatically fulfill the left- and right-canonical conditions. Therefore, the diagonal $\lambda$-matrices are identity matrices, and nontrivial diagonal matrices $\Lambda$ exist only on the multiplet level. Their eigenvalues $\Lambda_i$ determine the multiplet entanglement spectrum defined as
\begin{equation}
E_i=- \log(\rho_i) = -2\log(\Lambda_i),
\end{equation}
where $\rho_i = \Lambda_i^2$ is the reduced-density-matrix eigenvalue corresponding to each multiplet.

In order to illustrate the above concepts, let us now consider the spin-1 BLBQ model on a single chain [see Fig.~\ref{fig-sketch} (a) for the lattice geometry and (d) for corresponding MPS]. We use generalized boundary conditions on both ends of the MPS, in that the left (right) input bases of $A_1$ ($A_L$) can be specified as desired. The most natural choice in DMRG is to take the input basis to be a singlet $|S=0\rangle$, as usually done for open boundary conditions. In that case, however, the spin quantum number $S$ of the virtual bond multiplets would automatically be integer, as only integer $S$ results when adding two integer spins together. For this reason, SU(2) DMRG calculations with conventional open boundary conditions will never yield the half-integer bond (projective) representation of the SO(3) symmetry, but always a ``trivial" state without the expected (at least) two-fold degeneracy in each bond multiplet expected for the Haldane phase. 

On the other hand, the boundary can also be set up by taking both end bonds to be $|S=1/2\rangle$ doublets, instead of singlets $|S=0\rangle$.\cite{expl-sbc} Since then only half-integer multiplets appear in the virtual bonds, this always yields an ``SPT" state possessing doubly degenerate entanglement spectrum. In particular, for the spin-1 chain of Hamiltonian Eq.~(\ref{eq-bbq}), this choice of boundary condition produces an ``SPT" state for any $\theta \in [-\pi/2, \pi/4]$. However, this seemingly contradicts the well-established fact that Haldane phase is confined to $\theta \in (-\pi/4, \pi/4)$. In order to resolve this apparent paradox, we here also study a more general situation, where we input the direct sum $|S=0\rangle \oplus |S=1/2\rangle$ on the two boundary bonds. This gives rise to the possibility of both integer and half-integer multiplets on the bonds, and allows us to do actually parallel DMRG calculations in two independent symmetry sections, i.e., integer and half-integer bond spaces. 

We thus adopt the following \emph{protocol algorithm for determining the bond representations}: we input both integer and half-integer multiplets on the boundary virtual bonds, and perform several DMRG sweeps back and forth. In the presence of state space truncation along the bonds, depending on the Hamiltonian parameters, the system will eventually converge to the half-integer projective representation or the integer linear representation of SO(3), thus telling the SPT phase from a trivial one. 

Two typical ``multiplet entanglement spectra" selected through DMRG sweeps, and calculated using $|S=0\rangle \oplus |S=1/2\rangle$ boundary states, are shown in Fig.~\ref{fig-ent-spec} (a). Here each data point represents a multiplet, in contrast to the traditional state entanglement spectrum, where each data point corresponds to an individual state. $\theta=0$ corresponds to the conventional Heisenberg model, whose ground state belongs to the Haldane phase. The converged multiplet spectrum obtained is shown using asterisks: all points in the spectrum correspond to half-integer quantum numbers $S$, and each asterisk with quantum number $S$ represents $2S+1$ (an even number) degenerate U(1) states, as expected for an SPT phase. On the other hand, the system with $\theta=-\pi/2$ is in the dimerized phase, a trivial gapped phase. Its SU(2) multiplet spectrum is plotted using open circles in Fig.~\ref{fig-ent-spec} (a). In contrast to the $\theta=0$ case, the circles are all located at integer $S$, as expected for a trivial (non-SPT) phase.

In the protocol algorithm, where $|S=0 \rangle \oplus |S=1/2\rangle$ is used as auxiliary boundary state, DMRG allows the ``correct" bond representation to be found, as long as the system is not very close to the phase transition point. In the following, in order to compare the multiplet spectra between the integer and half-integer representations, we now change strategy and enforce the representation by specifying one of the two boundary state types on both ends of the chain, i.e., $|S=0\rangle$ ($|S=1/2 \rangle$) for integer (half-integer) representation.

In Fig.~\ref{fig-ent-spec} (b), we choose $\theta=0$ (corresponding for the Haldane phase), and compare the multiplet entanglement spectra $E^I_{i}$ (circle) and $E^H_i + \log(2)$ (asterisks), which are obtained by enforcing either integer or half-integer representations, respectively. The integer-spin multiplet spectrum evidently displays a two-fold degeneracy, whereas the half-integer-spin multiplet spectrum does not. Instead, we observe a one-to-one correspondence between each multiplet in $E^H_i + \log{(2)}$ and a pair of degenerate multiplets in $E^I_i$. The shift value $\log(2)$ is chosen because the two representations have different numbers of states with nonzero weights in their reduced density matrices. The nonzero individual states in the integer representation are twice as many as those in the half-integer one.

This different behavior of the degeneracies in the integer and half-integer multiplet entanglement spectra can be understood as follows: in the presence of space inversion symmetry, time reversal symmetry, or some $Z_2 \times Z_2$ rotational symmetry, etc., which protects the Haldane phase, it has been proven that $\Lambda_{Q_i} \otimes \lambda_{Q_i}$ has an even degeneracy of at least 2. \cite{Pollmann_1} Therefore in the Haldane phase, either $\Lambda_{Q_i}$ or $\lambda_{Q_i}$ should have even degeneracy. For the half-integer bond representation, the $Q_i$'s are half-integer and therefore the $\lambda_{Q_i}$'s are identity matrices with an even number of diagonal elements, implying that an even degeneracy appears in the CGC space; thus the $\Lambda_{Q_i}$ in the reduced multiplet space is not necessarily two-fold degenerate, which explains the absence of degeneracies in the multiplet spectrum $E_i^H$ (asterisks). On the other hand, for integer bond representations, the $\lambda_{Q_i}$'s are identity matrices of odd-number rank, therefore an even degeneracy must instead appear on the multiplet level, which explains the two-fold degeneracy obtained in $E_i^I$ (circles). This difference between integer and half-integer representations has an important consequence in the entanglement entropy, which will be discussed in the next section.

To summarize, the lesson learnt from Fig. \ref{fig-ent-spec} is as follows. In Fig. \ref{fig-ent-spec} (a) we showed that, if mixed boundary $|S=0 \rangle \oplus |S=1/2\rangle$ is adopted, DMRG sweep can select the half-integer-spin representation in the Haldane phase and integer-spin representation in the trivial phase. Fig. \ref{fig-ent-spec} (b) illustrates that if one studies the Haldane phase using auxiliary spin $|S=0\rangle$ or $|S=1/2\rangle$ on the external bond, respectively, then the general requirement of having an entanglement spectrum of even degeneracy is satisfied by having the multiplet spectrum being degenerate or non-degenerate for the case of integer-spin or half-integer-spin representation, respectively.

\begin{figure}[tbp]
\includegraphics[angle=0,width=0.95\linewidth]{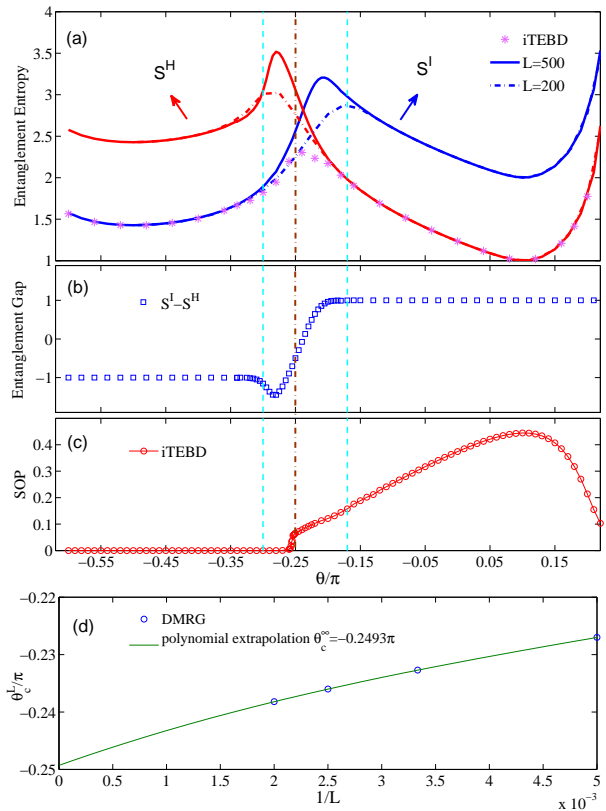}
\caption{(Color online) Integer and half-integer entanglement entropies, $S^I$ and $S^H$, of the spin-1 BLBQ model, for $L=200$ (dash-dotted lines) and $L=500$ (solid lines). Results for different system sizes coincide for $\theta$-values far from the critical point at $\theta_c=-\pi/4$ (vertical dash-dotted line), but differ in the intermediate region between the two vertical dashed lines. $S^I$ and $S^H$ cross at a ``pseudo-transition" point $\theta_c^{L}$, which moves towards the critical point as the system size is increased (shown in panel (d), the extrapolated point is very close to the true critical one). In the above calculations, 400 multiplets (about 1600 individual states) have been retained, the truncation errors are of the order $10^{-10}$ around critical point, and are negligible ($10^{-14}$ or less) for the rest parameters. The entanglement entropies are evaluated at the center of the chain. Panel (a) also shows the entanglement entropy obtained by the iTEBD algorithm\cite{iTEBD} (asterisks), which favors the minimally entangled states, and always follows the lower entanglement entropies. (b) The entanglement gap $G=S^I-S^H$, which equals $\pm1$ in the SPT phase and the trivial phase respectively. The dashed vertical lines mark the intermediate region, where $G$ is not a constant owing to finite size effects. Panel (c) shows the string order parameter (SOP $O_z$) of Eq.~(\ref{eq-SOP}), obtained by iTEBD calculations, which retain up to 200 states.}
\label{fig-chain}
\end{figure}

\section{Entanglement Gap and Symmetry-Protected Topological Phase}
During the DMRG sweeps in the protocol using $|S=0 \rangle \oplus |S=1/2\rangle$ as boundary, as long as the doublet $|S=1/2\rangle$ is not physically coupled to the bulk (the coupling strength between the auxiliary boundary spin-1/2 and the spin-1 chain can be set to be very weak or even turned off), the integer and half-integer symmetry sectors have exactly the same ground-state energy. Therefore, the energy is irrelevant in selecting the symmetry sector in the protocol algorithm. Instead, since the two-site update scheme of DMRG is adopted during the sweeps, the truncation and hence the entanglement entropy is important in selecting the symmetry sector.

In order to uncover this mechanism underlying the protocol algorithm, we now study the bipartite entanglement entropies in the integer and half-integer symmetry sectors, respectively, by enforcing different boundary states. The entanglement entropies are defined as 
\begin{equation}
S^{X} = - \sum_{Q} {\rm{Tr}_{Q}} [(\rho^X_{Q} \otimes D^X_{Q}) \log_2(\rho^X_{Q} \otimes D^X_{Q})], 
\label{eq-ent-ent}
\end{equation}
where $X= H$ or $I$ for half-integer or integer representations, and the difference $G=S^I - S^H$ will be called the ``entanglement gap". In Eq. (\ref{eq-ent-ent}), $\rho^X$ is the reduced density matrix on the multiplet level. It is block-diagonal, with blocks $\rho^X_{Q}$ labeled by $Q$ and matrix elements $(\rho^X_{Q})_{n,n'}$. $D^X$ is an identity matrix, with matrix elements $(D^X_Q)_{Q_z,Q'_z} = \delta_{Q_z,Q'_z}$ whose trace thus equals the inner dimension of each multiplet.. Consequently, ${\rm{Tr}}_Q[\cdot]$ refers to the trace over both, the multiplet index $n$ as well as as the internal multiplet space $Q_z$ of a given symmetry sector $Q$. Note that the logarithm to base 2 ($\log_2$) is adopted in evaluating the entanglement entropy. $\rho^X$ and $D^X$ are readily obtained from DMRG simulations. We note the SU(2) multiplet language used to formulate
Eq. (6) for the von Neumann entropy can easily be applied to also calculate the Renyi entropy. Very recently, the latter has been employed  to study the local differential convertibility and thereby probe the SPT phase.\cite{Cui} Though we here focus only on the von Neumann entropy, our analysis can be be generalized straightforwardly to the Renyi entropy.

In Fig.~\ref{fig-chain}, $S^I$ and $S^H$ of the spin-1 BLBQ chain [Eq.~(\ref{eq-bbq})] are plotted in (a), and the entanglement gap $G$ is shown in (b). For the Haldane phase ($-\pi/4<\theta<\pi/4$ in Fig.~\ref{fig-chain}) we find $S^I > S^H$, thus the half-integer bond representation has lower entanglement than the integer one, although the ground-state energies in both representations are the same. On the other hand, for the dimerized phase ($-0.6 \pi<\theta<-\pi/4$ in Fig.~\ref{fig-chain}) we find $S^I < S^H$. This observation explains why the protocol using $|S=0\rangle \oplus |S=1/2\rangle $ on end bonds employed for Fig. \ref{fig-ent-spec}~(a) succeeded in selecting the ``correct" bond representations: DMRG always favors lower entanglement, and the representation (integer or half-integer) with higher entanglement would be discarded by truncations during the sweeps.

Another interesting observation is that the entanglement gap $G = S^I-S^H$ is found to be a constant +1 ($-1$) in the SPT (trivial) phase. It is rather robust and almost independent of different Hamiltonian parameters and system sizes, except for the intermediate region near the critical point, where the finite-size effects become significant. This region is marked by vertical dashed lines in Fig.~\ref{fig-chain}. The entanglement curves cross within this region, and the crossing point moves to the true critical point $\theta_c = -\pi/4$, the exactly soluble Takhtajan-Bubujian point, \cite{TB} with increasing system sizes. 

The value $G = \pm1$ actually originates from the different topology of the SU(2) and SO(3) groups, and hence can be regarded as a topological invariant in each phase. In order to understand this, let us again consider the exactly solvable AKLT model with $\theta=\arctan(1/3)$. The reduced tensor at the multiplet level is $A_{S=1/2, S=1/2}^{S=1}=1$, a simple tensor with bond dimension 1, i.e.,  a scalar number. The corresponding CGC tensor is $C(S=1/2, S=1 | S=1/2)$, which combines a spin doublet with a triplet into an output spin doublet. 

The corresponding reduced density matrix of half-infinite AKLT chain is a $2\times 2$ diagonal matrix, $ \begin{pmatrix} 1/2 & & 0 \\ 0 & & 1/2 \end{pmatrix}$, fully encoded in the CGC space only, and resulting in an entanglement entropy $S^H=2[-1/2\log_2(1/2)]=1$. However, for the integer bond representation, we instead have a 2 $\times$ 2 diagonal matrix 
\begin{equation}
\begin{pmatrix} 
A_{S=0, S=0}^{S=1} = 1/4 & & 0 \\ 0 & & A_{S=1, S=1}^{S=1} = 1/4 
\end{pmatrix}  
\end{equation}
in the reduced multiplet space. The two degenerate multiplets contain 4 degenerate states in total, and the full reduced density matrix is a $4 \times 4$ diagonal matrix with all elements $1/4$. The entanglement entropy is $S^I=4[-1/4\log_2(1/4)]=2$, which is larger than corresponding $S^H$ and the gap $G = S^I- S^H = 1$. 

Next, we consider a generic state in the SPT phase away from the special AKLT point. As shown in Fig.~\ref{fig-ent-spec}, there exists a one-to-one correspondence between one $\tilde{S}=(2n+1)/2$ multiplet in the half-integer sector and one pair of degenerate multiplets with $S=n$ and $n+1$ in the integer sector ($n = 0, 1, 2, ...$). For the latter, the degeneracy on the multiplet level cannot be trivially lifted owing to the protection of the symmetry. Consequently, this multiplet degeneracy enhances the entanglement entropies and opens an entanglement gap of $G = S^I- S^H = 1$, as shown in Fig.~\ref{fig-chain}. On the other hand, adopting integer virtual bonds would preferably lower the entropy by 1 for the trivial dimer phase. As shown in Fig.~\ref{fig-chain}, in this case entanglement gap is $G=-1$. 
 
Fig.~\ref{fig-chain}(c) presents the nonlocal string order parameter obtained by iTEBD calculations; it is nonzero in the Haldane phase and vanishes in the trivial phase. The comparison of our entanglement entropy results with the SOP data validates that $G$ can be used to distinguish SPT phase from the trivial one.

Lastly, we remark that the results in Fig.~\ref{fig-chain} were obtained by evaluating finite-size systems. When the system is close to the critical point, the entanglement entropies $S^I$ and $S^H$ are shown to cross each other. In Figs.~\ref{fig-chain}(a), the lower values of these two entropy curves can be regarded as giving the ``true" entanglement entropies. The combined curve shows a sharp peak, which is missed when considering either $S^I$ or $S^H$ alone. In Fig.~\ref{fig-chain}(a), the results obtained by iTEBD are also shown, which always favor the low entanglement curves. The iTEBD data coincide with the SU(2) DMRG results (except for the region near the critical point), which validates our arguments above. The crossing point of the $S^I$ and $S^H$ curves (as the peak of the low entanglement curve) can be viewed as a ``pseudo-transition" point. As the system size increases, the pseudo-transition point approaches the true critical point $\theta_c = -\pi/4$ [see Fig. \ref{fig-chain}(d)]. In the thermodynamic limit, the gap $G$ are supposed to show a jump between 1 and $-1$ just at the critical point, and the peaks of the entanglement entropies are expected to diverge.
 
\section{Spin-1 Heisenberg Tube Model}

\begin{figure}[tbp]
\includegraphics[angle=0, width=0.95\linewidth]{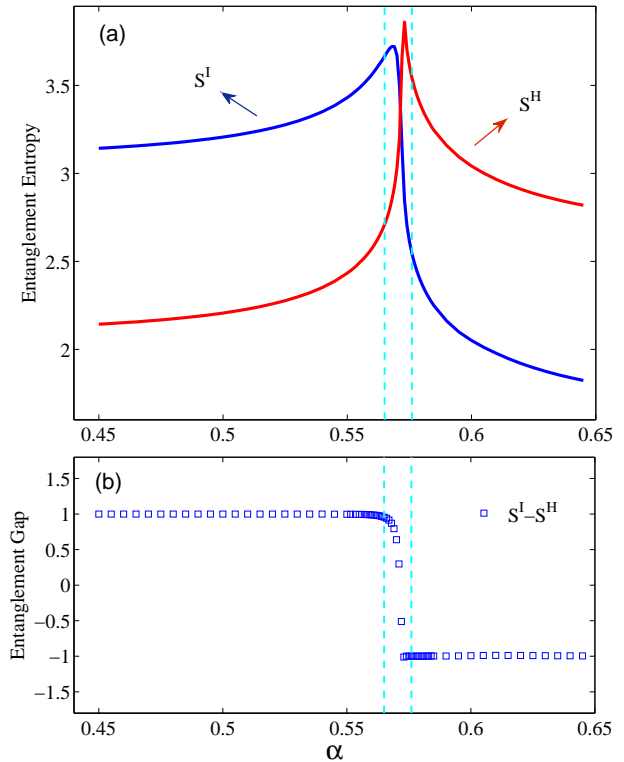}
\caption{(Color online)  (a) Integer and half-integer entanglement entropies $S^I$ and $S^H$ for the spin-1 tube model. The critical point estimated from their crossing point is $\alpha_c = 0.571(1)$. (b) Entanglement gap $G = S^I-S^H$. $G=1$ when $\alpha<\alpha_c$, identifying the existence of an SPT phase, and $G=-1$ when $\alpha<\alpha_c$, corresponding to a trivial phase. The system size is $3 \times 100$, 400 multiplets are reserved, which lead to maximum truncation error of $10^{-8}$ (at the critical point).}
\label{fig-tube}
\end{figure}

\begin{figure}[tbp]
\includegraphics[angle=0, width=0.9\linewidth]{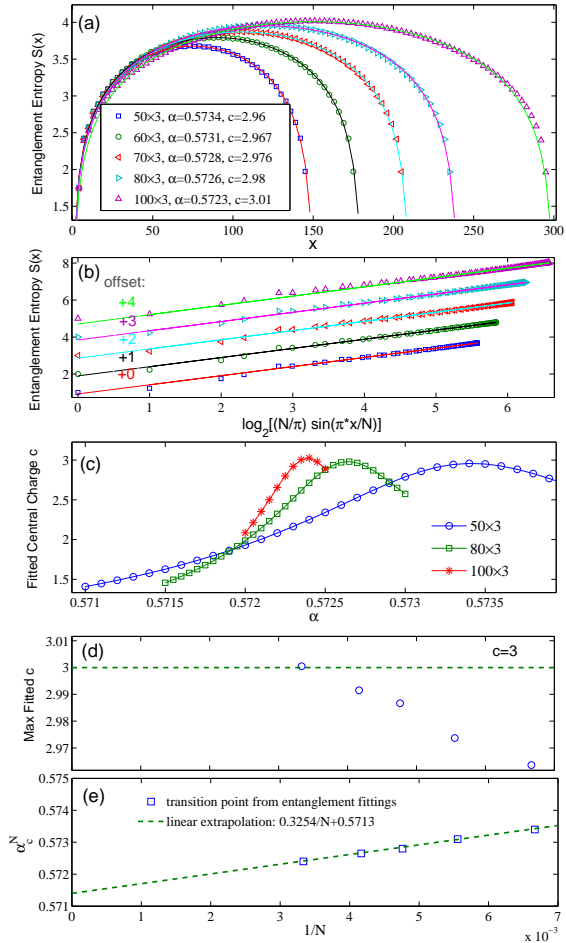}
\caption{(Color online) Analysis of the block entanglement entropy, $S(x)$, of the ground state of the spin-1 three-leg tube Heisenberg model, in the vicinity of $\alpha_c$. (a) and (b) show the entanglement entropy between boundary block of length x and the rest of the system, for several different system sizes and $\alpha$-values, on (a) a linear scale and (b) a log(sin) scale on the horizontal axis. Curves are vertically offset by 1 unit for clarity. Here we show the data on one of the three sublattices in tube model, which contains entanglement entropies cut at the $i$-th bond [$\rm{mod}(i,3)=1$]; the other two curves give the same fitting results and are not present here. The conformal central charge is determined as $c \simeq 3$, (c) shows how the fitted $c$'s vary with $\alpha$, for three fixed system sizes. (d) and (e) show, respectively, the maximal $c$-values and corresponding $\alpha$-values obtained for 5 different $N$-values. The system size ranges from $N=50\times3$ to $100\times3$ ($N$ is the total site number), and up to 450 bond multiplets are reserved in the calculations. A half-integer bond representation was adopted in the calculations; the fittings of integer-representation entanglement entropies lead to the same conclusion.}
\label{fig-fitc}
\end{figure}

 \begin{figure}[tbp]
\includegraphics[angle=0, width=1\linewidth]{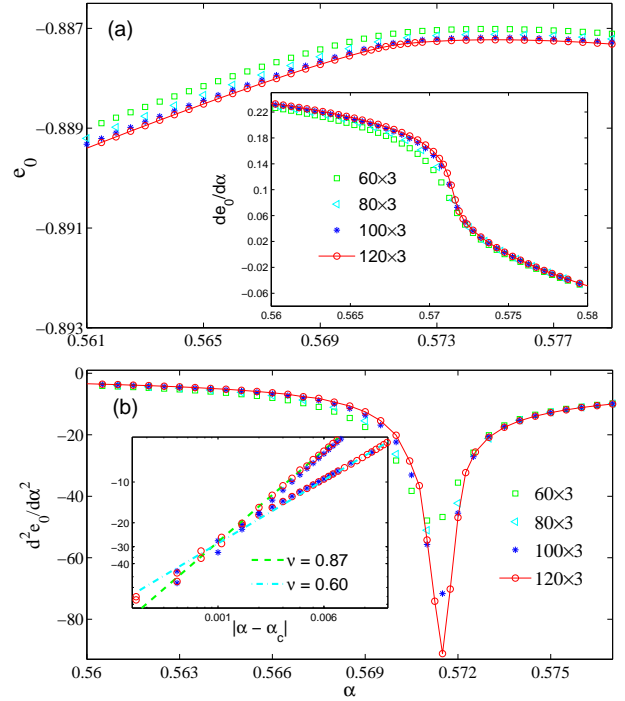}
\caption{(Color online) (a) The ground state energy per site $e_0$ of a spin-1 tube versus the coupling ratio $\alpha$. The system size varies from 60$\times$3 to 120$\times$3. For the largest size $120 \times 3$, 500 SU(2) multiplets ($\approx 2000$ equivalent U(1) states) are retained in the calculations, truncation errors are less than $10^{-9}$. The inset shows the first-order derivatives of energies $de_o/d\alpha$, which are substantially converged with different system sizes, and are shown clearly to be continuous through the critical point. (b) The second-order derivative $d^2 e_0/ d \alpha^2$, which shows a diverging peak at $\alpha_c = 0.5715(5)$. The inset in (b) shows $d^2e_0/d\alpha^2$ in the vicinity of critical point on a log-log scale. The data points fall into two linear lines (except for the points very close to the critical point $\alpha_c$, owing to the finite-size effects near the critical point), which implies algebraic divergence. The dashed lines in the inset are fits to the form $d^2 e_0/d \alpha^2 \propto (\alpha - \alpha_c)^{-\nu}$, with $\nu \simeq 0.87$ and $0.6$, approaching critical point from left and right sides, respectively.}
\label{fig-eng-curv}
\end{figure}

In this section, we study the SPT phase in a spin-1 tube model. This model has been studied by Charrier et. al. in Ref. \onlinecite{Charrier}. Following their conventions, schematically depicted in Fig.~\ref{fig-sketch} (c), the Hamiltonian is given by: 
\begin{eqnarray}
& H_{tube} & = H_L + H_R, \notag \\
& H_L & = J_L \sum_{i, a = \{1,2,3\}} \bold{S}_{i, a} \bold{S}_{i+1, a}, \notag \\
& H_R & = J_R \sum_{i} (\bold{S}_{i,1} \bold{S}_{i,2} + \bold{S}_{i, 2} \bold{S}_{i,3} + \alpha \bold{S}_{i, 1} \bold{S}_{i,3}).
\end{eqnarray} 
$H_L$ and $H_R$ are the intra- and inter-chain coupling terms, respectively. In Ref. \onlinecite{Charrier}, the authors found a Haldane phase existing for $0< \alpha <0.57$, with $J_L=0.1$ and $J_R=1$, where each triangle contains an effective spin-1. For $0.57<\alpha<1.5$, they found a trivial disordered phase, with each isosceles triangle carries an effective spin-0, leading to a spin-0 chain (note that the combined product space 1 $\otimes$ 1 $\otimes$ 1 allows for exactly one spin-0 singlet). At the critical point $\alpha_c$, the system undergoes a quantum phase transition between the Haldane and the trivial phase. For $0< J_L/J_R < 0.65$, there are still phase transitions separating two phases, but at different $\alpha_c$; if $J_L/J_R > 0.65$, no phase transition occurs because the trivial phase no longer exists. \cite{Charrier}

Next, we revisit this model using SU(2) DMRG calculations, and study it by evaluating the entanglement entropies $S^I$ and $S^H$. In Fig.~\ref{fig-tube}, the entropies $S^I$ and $S^H$ intersect at $\alpha_c\approx 0.571(1)$. For $\alpha < \alpha_c$, $G = S^I - S^H = 1$, the half-integer representation has lower entanglement. The dominating bond multiplets are doublets ($S$=1/2), and the system is in an SPT (Haldane) phase. In contrast, for $\alpha > \alpha_c$, $G = S^I - S^H = -1$, the ground state favor integer bond representations. The energy results show that the energy per triangle is uniform along the leg direction, without any translational symmetry breaking. The leading bond multiplet in the entanglement spectrum is found to be a singlet ($S$=0), and the system is in a trivial disordered phase. In addition, we remark that the proper definition of a SOP in this spin-1 tube has been discussed by the authors in Ref.~\onlinecite{Charrier}. The SPT phase that we have here identified by entanglement entropy, indeed also possesses a nonzero SOP.

Compared with the spin-1 BLBQ model, finite-size effects are much less significant in the spin tube model. For a system size of $100 \times 3$, the values at which the peaks of integer and half-integer entropies occur lie quite close together. By combining $S^I$ of $\alpha>\alpha_c$ and $S^H$ of $\alpha<\alpha_c$, we can see a very sharp peak in the joint low entanglement curve, which suggests a second-order quantum phase transition.

Next, we address the order of the phase transition in more details by checking the criticality at $\alpha_c$. The block entanglement entropy of size $x$ can be fitted with the following form:
\begin{equation}
S(x) = \frac{c}{6} \log_2[ \frac{N}{\pi} \sin(\pi \frac{x}{N})] + \rm{const.}, 
\label{eq-CC}
\end{equation}
where $N$ is the total number of sites. $N=3L$ for the tube of length $L$. This is the Cardy-Calabrese formula \cite{Holzhey,Vidal_2,Calabrese} with open boundary condition, showing that the block entanglement entropy has a logarithmic correction to the entanglement area law at the critical point.\cite{Eisert} $c$ is the conformal central charge, which characterizes the criticality. The fitting results are shown in Fig.~\ref{fig-fitc}, which strongly suggests that the transition point is critical or very close to some gapless point (quasi-critical). The central charge obtained from the fits is $c\simeq3$. 

By the DMRG ordering of sites into one linear sequence, the 3-leg tube has three different sublattices (and hence three kinds of bonds), two of which are equivalent. Therefore, when cutting the systems in different ways, we can get three block entanglement entropy curves, one of which is shown in Fig. \ref{fig-fitc}. The fittings of the other two curves lead to the same results. Fig. \ref{fig-fitc}(a) and (b) show fits for 5 different system sizes and $\alpha$-values. Fig. \ref{fig-fitc}(c) shows that the $c$-values obtained from each fit exhibit, for given tube with total site number $N$, a clear maximum as function of $\alpha$. This maximal value (located at $\alpha_c^N$) can be regarded as the best estimation of $c$. Note, the system is most close to critical at $\alpha_c^N$, and away from critical when $\alpha < \alpha_c^N$ and $\alpha > \alpha_c^N$; Cardy-Calabrese formula (Eq. \ref{eq-CC}) gradually loses its legitimacy in the latter case, and fitted value of $c$ is reduced away from $\alpha_c^N$. Collecting these maximal points, in Figs. \ref{fig-fitc}(d) and (e) we plot, respectively, how the fitted $c$'s and estimated transition points $\alpha_c^N$'s vary with different system sizes (from $50\times3$ to $100\times3$). The fitted $c$'s (estimated transition points from entanglement) tend towards 3 (critical point estimated from energy derivatives) when $N$ is increased. Moreover, in the fits, we follow the same strategy as in Refs. \onlinecite{Laeuchli, Nishimoto} and fit the central charge in the central region of the chain. Typically we omit 10 to 20 sites (depending on the total system sizes) from both ends, and take $c$ to be the limiting value obtained when increasing the omitted site number.

The ground-state energy curves and their derivatives with respect to $\alpha$ are presented in Fig.~\ref{fig-eng-curv}. The energy per site is defined as $e_o = E_{\rm{tot}}/N$, where $E_{\rm{tot}}$ is the total energy and $N$ is the number of sites. The first-order derivatives of energies do not show any discontinuities at the transition point, but the second-order derivatives have very sharp peaks at $\alpha_c$. In the inset of Fig.~\ref{fig-eng-curv} (b), we also plot $d^2 e_0/d \alpha^2$ on a log-log scale. The observed power law behavior implies the algebraic divergence of $d^2 e_0/d \alpha^2$ approaching $\alpha_c = 0.5715(5)$, i.e., $d^2 e_0/d \alpha^2 \propto (\alpha-\alpha_c)^{-\nu}$. The exponent $\nu$ has two different values, depending from which side $\alpha_c$ is approached. Both, though, are less than 1, which implies that $de_0/d\alpha$ maintains a smooth behavior at $\alpha_c$. Therefore, the results of entanglement entropies, block entropy fittings, along with the energy derivatives, all support the conclusion that there is a continuous phase transition at $\alpha_c$. This contradicts the conclusion in Ref. \onlinecite{Charrier}, where the transition is argued to be of weakly first-order.

In order to thoroughly clarify the transition order, more detailed studies of the correlation functions and excitation gaps are needed, which we leave as future studies. The parameters could also be tuned (say, take $J_R/J_L$ different from $0.1$ studied above) and investigate the nature of the phase transition; or introduce some other parameters in the Hamiltonian (say bilinear-biquadratic parameter $\theta$) and inspect the transition along some other paths in the parameter space. We have done some preliminary calculations along these lines (not shown in this paper), which reinforce the conclusion of a second-order phase transition.
\section{Absence of Symmetry-Protected Topological Phase in Spin-1 Heisenberg Ladder}

\begin{figure}[tbp]
\includegraphics[angle=0, width=0.95\linewidth]{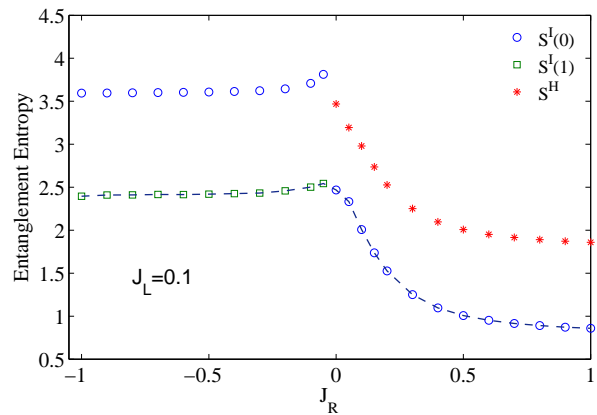}
\caption{(Color online) The entanglement entropies of spin-1 two-leg ladder system with system size $80 \times 2$. 200 multiplets are retained in the calculations, and the maximum truncation errors $ \approx 10^{-8}$. (a) $S^I$ and $S^H$ represent integer and half-integer entropies, respectively. $S^I(0)$ or  $S^I(1)$ means that $|S=0 \rangle$ or $|S=1 \rangle$ dominates in the multiplet spectrum, respectively. The dashed line is a guide for the eye.}
\label{fig-ladder}
\end{figure}

Lastly, let us consider the spin-1 two-leg ladder model, 
\begin{equation}
H = J_L \sum_{i, a=\{1,2\}} S_{i, a} S_{i+1, a} + J_R \sum_{i} S_{i, 1} S_{i,2}.
\label{eq-ladder}
\end{equation}
There are two kinds of couplings in this model [see Fig.~\ref{fig-sketch} (b)], $J_L$ along the chain direction and $J_R$ on the rungs. In Fig.~\ref{fig-ladder}, the entropies $S^I$ and $S^H$ are plotted. Two versions of $S^I$ are shown, $S^I(0)$ and $S^I(1)$, both obtained with integer bond representations, but with different leading (lowest) multiplets in the entanglement spectrum: $|S=0\rangle$ for $S^I(0)$ and $|S=1 \rangle$ for $S^I(1)$. The latter can be obtained by attaching auxilliary spin-1's on both ends in our SU(2) DMRG.

For $J_R>0$, $G = S^I(0)-S^H>0$ in Fig.~\ref{fig-ladder}, thus the ground state favors integer-spin representation, verifying the triviality of the ground state. Indeed, for the limiting case $J_R/J_L \to \infty$, the ground state is a simple direct product of rung singlets. On the other side, for $J_R<0$ the system is in the same phase as the spin-2 antiferromagnetic Heisenberg chain (reached in the limiting case $J_R/J_L \to -\infty$). Fig.~\ref{fig-ladder} shows that the ground states in this region also favor integer representations. However, the lowest multiplet in the entanglement spectrum is the spin triplet $|S=1\rangle$, rather than the singlet $|S=0 \rangle$, consistent with the results of Ref. \onlinecite{Pollmann_2}. The two low-entanglement curves from the $S=0$ and $S=1$ symmetry sectors together form a smooth line in Fig.~\ref{fig-ladder} (a) (indicated by a dashed line), which represents the ``true" entanglement entropy of the system. 

No sign of criticality can be seen from the entanglement entropies, and it is hence believed that only one disordered phase exists in the spin-1 Heisenberg ladder model. Our observation is in agreement with the conclusion in Ref.\onlinecite{Todo}, that the model does not undergo any phase transition from $J_R<0$ to $J_R>0$. The fact that there does not exist an SPT phase in the spin-1 Heisenberg ladder model studied above can be ascribed to the triviality of the standard $S=2$ AKLT states, which can be adiabatically connected to the topologically trivial state without any phase transition. \cite{Pollmann_2, Oshikawa, Tonegawa} The triviality of the standard $S=2$ AKLT state can be also be intuitively understood as follows: it has \textit{two} valence bonds (corresponding to two virtual spin-1/2) living on each geometric bond, since these two virtual spin-1/2 couple to either spin 0 or 1, the total spin forms integer-spin representations of SO(3) on the geometric bond, leading to a conclusion of a topologically trivial phase. This argument also applies to the spin-1 Heisenberg ladder studied above (especially when $J_R<0$). The bond states are more complicated for the general two-leg Heisenberg ladder model, nevertheless, they form integer representations of SO(3) and the corresponding groundstate belongs to a trivial phase.

\section{Conclusion}
We have proposed a novel way to identify SPT phases in one dimension by evaluating entanglement entropies. With SU(2) DMRG method, we can keep track of the bond multiplets, and readily tell half-integer-spin projective representation from integer-spin ones by checking the multiplet entanglement spectrum introduced in this paper. In addition, we have shown that auxiliary boundary spins attached on both ends of the chain can be used to control the bond representations; this significantly changes the entanglement entropies in the bulk, depending on the topological properties of the phase.

In the SPT phase, we showed that a two-fold degeneracy for the overall entanglement spectrum appears either in the reduced multiplet space or in the CGC space, depending on whether the integer or half-integer bond representations are adopted, respectively. In the latter case, the two-fold degeneracy occurs in CGC space, which reduces the entanglement entropy $S^H$ relative to $S^I$ (entanglement gap $G=1$), providing a practical criterion for identifying SPT phases. The existence of an entanglement entropy gap also allows us to automatically select the ``correct" representation (integer or half-integer) through DMRG sweeps, which always favor low entanglement representation. The entanglement gap closes at the critical point, which can be used to detect the quantum phase transitions.

Several 1D and quasi-1D systems have been studied in this work; the SPT phase in the spin-1 chain and the spin-1 tube model are successfully identified by evaluating the entanglement entropies. For the spin-1 tube model, the numerical results indicate that the phase transition between the SPT phase and the trivial phase is a continuous one. The fact that the two-leg spin-1 Heisenberg ladder has no SPT phase for any $J_R$ is also validated by our entropy results.

\section{Acknowledgement}
WL would like to thank H.-H. Tu and T. Quella for helpful discussions on symmetry-protected topological order and the projective representations of symmetry groups. WL was also indebted to Shou-Shu Gong for useful discussions on the numerical results and DMRG techniques. 
This work was supported by the DFG through SFB-TR12, SFB631, the NIM Cluster of Excellence, and also WE4819/1-1 (AW).

\end{document}